# Electrical Transport of perpendicularly magnetized $L1_0$-MnGa and MnAl films

L. J. Zhu[1,2]* and J. H. Zhao[1]**

1. State Key Laboratory of Supperlattices and Microstructures, Institute of Semiconductors,
Chinese Academy of Sciences, P. O. Box 912, Beijing 100083, China
2. Cornell University, Ithaca, NY 14850, USA
*zhulijun0@gmail.com; **jhzhao@red.semi.ac.cn

Ferromagnetic films of $L1_0$-ordered MnGa and MnAl that exhibit giant perpendicular magnetic anisotropy and great controllability in the magnetism and structural disorders show promise not only in the applications in magnetic recording, permanent magnets and spintronics, but also in controllable studies of disorder-relevant electrical transport phenomena. In this article, we review the intriguing experimental observations of the orbital two-channel Kondo effect and anomalous Hall effect in $L1_0$-ordered MnGa and MnAl thin films with perpendicular magnetic anisotropy. We also give a perspective with regards to the future technological and fundamental applications of these perpendicularly magnetized Mn-based binary alloy films.

## 1. Introduction

Ferromagnetic films of perpendicularly magnetized Mn-binary alloys $Mn_xGa$ (MnGa) and $Mn_xAl$ (MnAl) with $L1_0$-ordering (Fig. 1) have been studied intensively in the past decades due to the technological and scientific interest. $L1_0$-MnGa (MnAl) alloys are theoretically expected to have perpendicular magnetocrystalline anisotropy (PMA) of 26 (15) Merg/cc,[1-5] saturation magnetization ($M_s$) of 845 (800) emu/cc, magnetic energy product $(BH)_{max}=(2\pi M_s)^2$ of 28 (12.64) MGOe,[6,7] respectively. Pump-probe time-resolved magneto-optic Kerr effect experiments revealed Gilbert damping constants ($\alpha$) as small as 0.008 in sputtered $L1_0$-MnGa films,[3] which are attributable to the weak spin-orbit coupling and the small density of states at Fermi level. $L1_0$-MnAl is also likely to have a small $\alpha$ because of the weak spin–orbit coupling of the Mn and Al elements despite the magnetization dynamics of $L1_0$-MnAl has so far remained unclear. The Curie temperature ($T_c$) of these films were reported to be high (Fig. 2(a)).[8] Recent studies have also demonstrated that high-quality single-crystalline $L1_0$-MnAl and MnGa films with fairly square perpendicular magnetization hysteresis loops can be epitaxially grown on semiconductors (Fig. 2(b)),[4,8] while the major magnetic properties including $M_s$, the coercivity, the PMA, and $(BH)_{max}$ can be remarkably engineered by controlling growth temperature ($T_s$), composition, and post-annealing.[4,8,9] The fascinating magnetic and structural properties make $L1_0$-MnGa and MnAl promising not only in their potential applications in ultrahigh-density perpendicular magnetic recording, economical permanent magnets, and spintronics, but also in controllable studies of disorder-relevant electrical transport phenomena.

Since there have been several review papers on the growth, the magnetic properties, and the application in magnetic tunneling junctions,[6,7,10] in this review, we mainly focus on the recent progress on the intriguing transport phenomena of $L1_0$-MnGa and MnAl films with controllable structural and electronic disorders, including the orbital two-channel Kondo (2CK) effect and the anomalous Hall effect (AHE). In the last part of this review, we will give an outlook on the potential applications of the $L1_0$-MnGa and MnAl films.

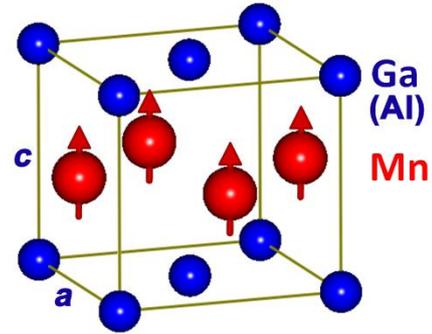

Fig. 1. The lattice structure of $L1_0$-ordered MnGa and MnAl.

## 2. Orbital two-channel Kondo effect

The overscreened 2CK effect displaying non-Fermi-liquid (NFL) physics has been of considerable scientific interest in recent years.[11-13] It may occur when a spin-1/2 impurity symmetrically couples to conduction electrons in two equal orbital channels via exchange interaction (spin 2CK),[11] or when a pseudospin-1/2 of two degenerate macroscopic charge states of a metallic island symmetrically couples to two conduction channels (charge 2CK),[13] or when a pseudospin-1/2 of structural two-level system (TLS, where an atom or atom group with small effective mass coherently tunnels between two nearby positions) equally couples to two spin channels of conduction electrons via resonant scattering (orbital 2CK, see Fig. 3(a)).[14,15] The 2CK effect is expected to have a unique low temperature ($T$) resistivity upturn ($\Delta\rho_{xx}$), which scales with $\ln T$ beyond the Kondo temperature ($T_K$), followed by an exotic NFL behavior ($\Delta\rho_{xx}\sim T^{1/2}$) as the consequence of two conduction electron spins screening the spin (pseudospin) impurity.[15-17] The charge 2CK effect and spin 2CK effect were clearly demonstrated and channel asymmetry effect was probed directly and quantitatively.[11-13] However, the orbital 2CK physics has been under heated debate with regards to its existence and stability with respect to the population imbalance of two spin channels due to the strong magnetic fields or ferromagnetic exchange splitting despite the intensive studies for almost 30 years.[11-17] As shown in Fig. 3(b), the orbital 2CK effect is expected to exhibit a three-regime resistivity upturn with a transition from the $T^{1/2}$ scaling to

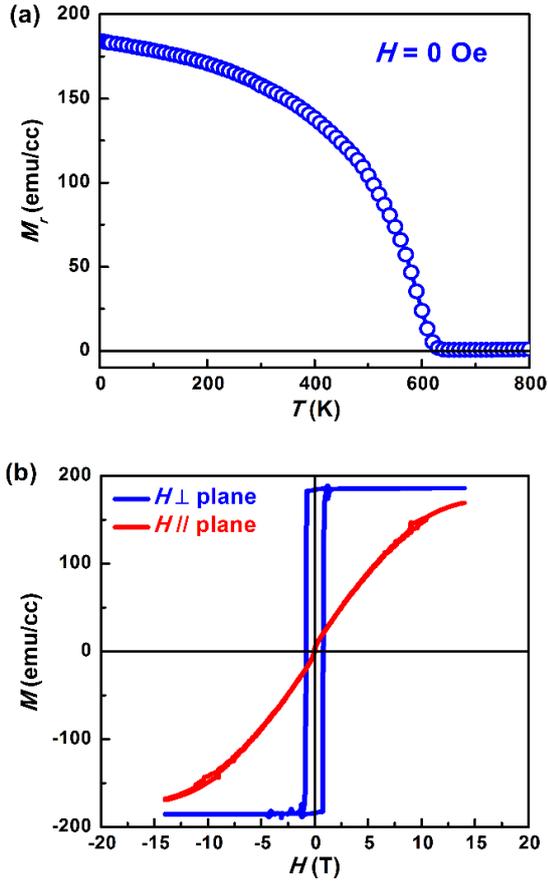

Fig. 2. (a) Temperature dependence of remnant magnetization and (b) the perpendicular and inplane magnetization hysteresis loops of a typical high-quality $L1_0$-MnGa film ($x = 1.4$) epitaxially grown on GaAs (001).

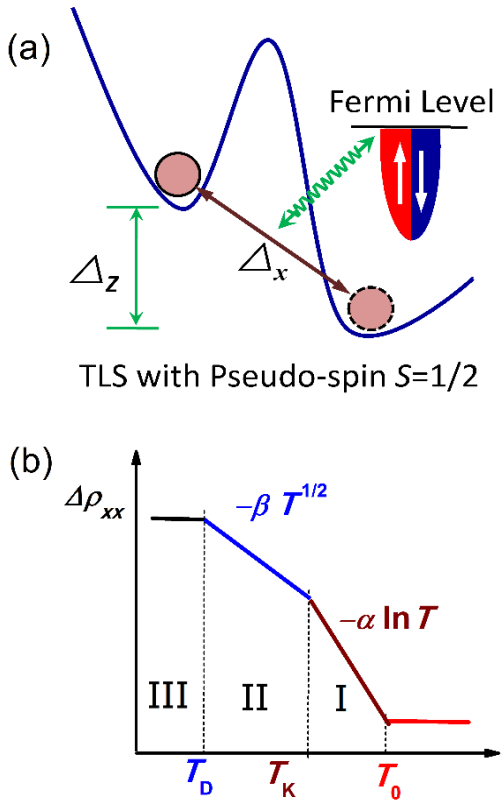

Fig. 3. (a) Schematic of 2CK effect induced by a two-level system (TLS) and (b) three-regime resistivity upturn of the orbital 2CK effect.

ln$T$ dependence beyond Kondo temperature ($T_K$) and a breakdown of the NFL behaviors below a characteristic temperature $T_D$. The orbital 2CK effect can be further experimentally evidenced by a robust independence of applied magnetic field and a close relevance to structural disorder.[16,17] Magnetic fields should not have any observable influence on the resonant levels, coupling strength, and thus the effect amplitude via changing the population balance of the two spin channels of the conduction electrons because the Zeeman splitting is negligibly small (~ 0.9 meV at $H$ = 8 T) in comparison to the width of energy band and $E_F$ of a host system (~10 eV), e.g. ferromagnetic $L1_0$-MnAl and MnGa.[1,5]

$L1_0$-MnAl and MnGa films are a good playground for the exploration of disorder-related phenomena, e.g. orbital 2CK effect. The magnetic and transport properties are strongly dependent on the structural disorders and may be conveniently tailored by varying the growth parameters, offering a convenient pathway to tune the relevant 2CK parameters. Furthermore, under perpendicular magnetic fields, the anisotropic magnetoresistance (MR) and spin disorder scattering-induced MR should be negligible in a film with large PMA because of the orthogonal magnetization-current relation and the large energy gap in spin wave excitation spectrum. This is highly amenable to study the intrinsic magnetic field dependence of an orbital 2CK effect. In this section, we review the recent observation of the orbital 2CK effect in $L1_0$-MnAl and MnGa films.

## 2.1 Orbital 2CK effect in disordered $L1_0$-MnAl films

In the past three decades, orbital 2CK effect was mainly studied in ballistic conductors of Cu and Ti point contacts fabricated by electron-beam lithography and diffusive conductors of ThAsSe and ZrAsSe glasses prepared by chemical vapor transport.[18-20] Recently, Zhu et al.[16] reported the first experimental evidence of TLS-induced orbital 2CK effect in a ferromagnetic system, $L1_0$-MnAl epitaxial films with strong PMA.

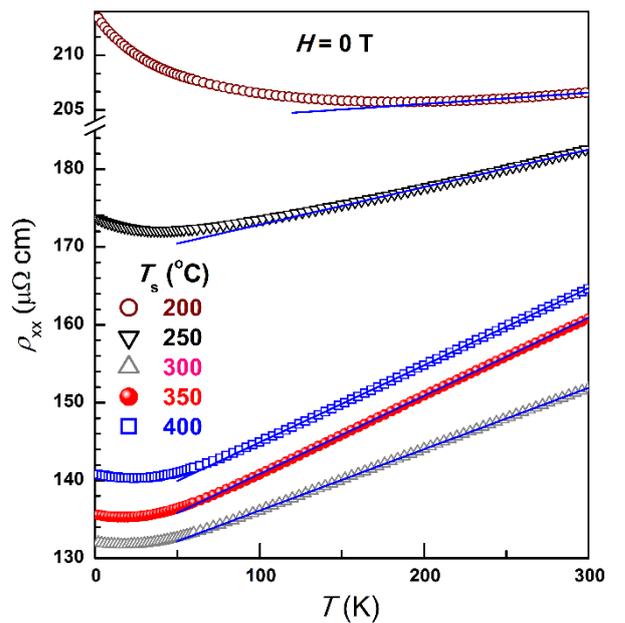

Fig. 4. Temperature dependence of $\rho_{xx}$ for the $L1_0$-MnAl films with different $T_s$. The solid lines stand for the best linear fits.

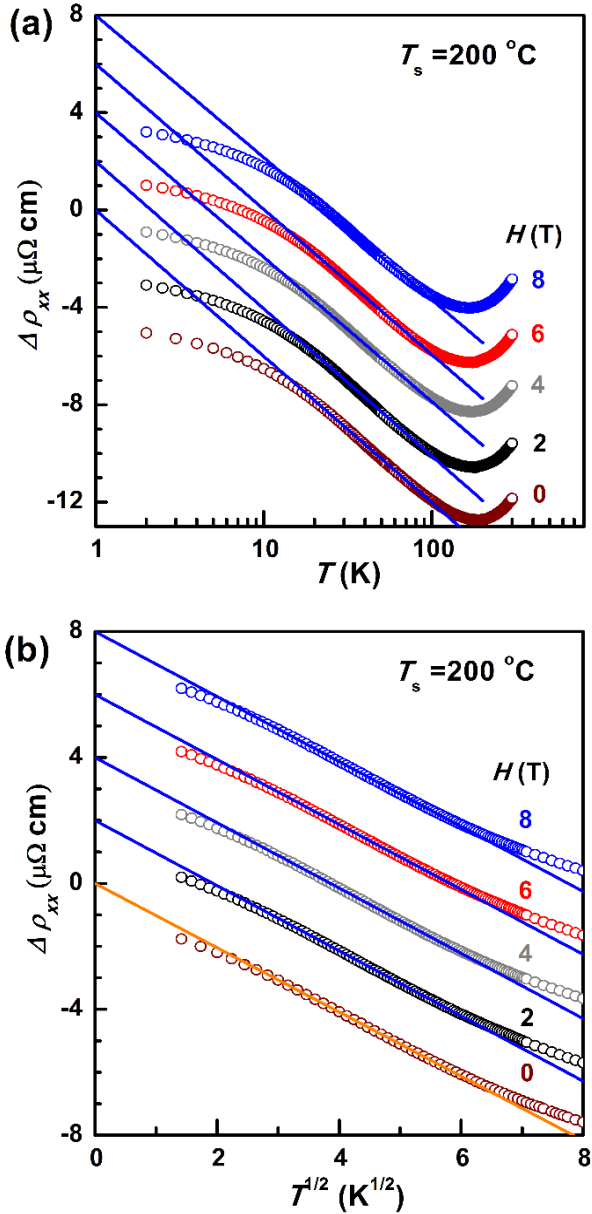

magnitudes (the slopes $\alpha = -d\rho_{xx}/d(\ln T)$ for $T_K < T < T_0$ and $\beta = -d\rho_{xx}/d(T^{1/2})$ for $T_D < T < T_K$) and characteristic temperatures ($T_0$, $T_K$, and $T_D$). This represents the first observation of all three theoretically-expected transport regimes from the orbital 2CK effect in the same samples.

As shown in the Fig. 6, the similar $T_s$ dependence of $\alpha$, $\beta$, and the density of active TLSs ($N_{TLS}$) agrees well with the variation of structural disorder evidenced by the intensity and the full width at half maximum of the $L1_0$-MnAl (002) peaks of x-ray diffraction patterns.[16] The close relevance of the resistivity upturn ($\alpha$ and $\beta$) to $N_{TLS}$ and structural imperfection further confirm the disorder nature of the orbital 2CK effect.

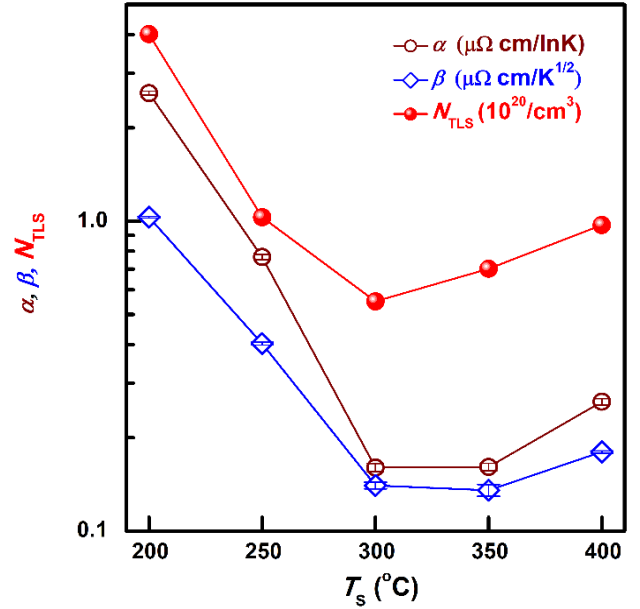

Fig. 6. $\alpha$, $\beta$ and $N_{TLS}$ plotted as a function of $T_s$ for $L1_0$-MnAl films.

## 2.2 Orbital 2CK effect in disordered $L1_0$-MnGa films

Similar to the case of $L1_0$-MnAl films, the structural disorder may also induce an orbital 2CK effect in $L1_0$-MnGa films.[17] Figure 7(a) and 7(b) show the $T$ dependence of $\rho_{xx}$ in a 30 nm thick $L1_0$-MnGa ($x = 0.94$) film with a small $M_s$ of ~100 emu/cm$^3$ at 300 K and a low $T_c$ of 366 K due to the significant structural disorder and the degraded crystalline quality. A low-$T$ resistivity upturn can be found for different $H$ of up to 8T, which first varies linearly with $\ln T$ below a temperature $T_0$ of ~25.5 K (Fig. 7(a)) and then crossover to a $T^{1/2}$ dependence (Fig. 7(b)) when $T$ drops below $T_K$ of 14.5 ±1.5 K. The high value of $T_K$ suggests strong Kondo coupling between the TLSs and conduction electrons via resonant scattering, in the case of which present theories expect an experimentally accessible orbital 2CK effect.[16] Here, both weak localization and electron-electron interaction (even if the diffusion channel is considered)[21] are also in qualitative contradiction to the apparent transition from the $\ln T$ scaling to the $T^{1/2}$ scaling at around $T_K$. The magnetic fields have no measurable influence on the scaling of $T$ dependence and the values of the slopes $\alpha$ and $\beta$ (Fig. 7(c)), strongly suggesting a nonmagnetic origin of the resistivity upturn observed in $L1_0$-MnGa. Specifically, there is no measurable change in $T_K$ under different $H$ (Figs. 7(a) and 7(b)), suggesting a negligible effect of $H$

Fig. 5. (a) Semilog plot of $\Delta\rho_{xx}$ versus $T$ and (b) $\Delta\rho_{xx}$ versus $T^{1/2}$ under different perpendicular magnetic fields for for $L1_0$-MnAl films ($T_s$=200 °C). For clarity, the curves in nonzero fields are artificially shifted by steps of 2 μΩ cm in (a) and (b).

Figure 4 shows the $T$ dependence of the zero-field longitudinal resistivity ($\rho_{xx}$) of a series of 30 nm thick $L1_0$-MnAl single-crystalline films grown at 200, 250, 300, 350, and 400 °C, respectively. For each film, $\rho_{xx}$ increases linearly with $T$ at high temperatures due to increasing phonon scattering, while the low-$T$ resistivity upturn most likely arises from the TLS-induced orbital 2CK effect.[16] As an example, Figures 5(a) and 5(b) show the $T$ dependence of resistivity variation at different $H$ for the $L1_0$-MnAl films, which shows distinct signatures associated with the TLS-induced 2CK effect. The low-$T$ resistivity upturn shows a $\ln T$ dependence below a temperature $T_0$ (Fig. 5(a)) with a clear transition to NFL behavior signified by a $T^{1/2}$ dependence when $T$ drops below $T_K$ and deviation from it upon further cooling to below $T_D$ (Fig. 5(b)). The scaling of the resistivity increase shows robust independence of strong applied magnetic fields of up to 8 T with regards to the on the Kondo coupling strength, tunneling symmetry, and barrier height of the TLSs. However, the $L1_0$-MnGa epitaxial film does not show any sign of a breakdown of the NFL behavior by a magnetic field of up to 8 T in the temperature range from 2 to 300 K, which suggests both a negligible influence of the applied magnetic fields on the population balance of the two spin channels and the robustness of the 2CK physics to a slight population imbalance. These observations provide strong evidence for the orbital 2CK effect being induced by TLSs originating from nonmagnetic impurities.

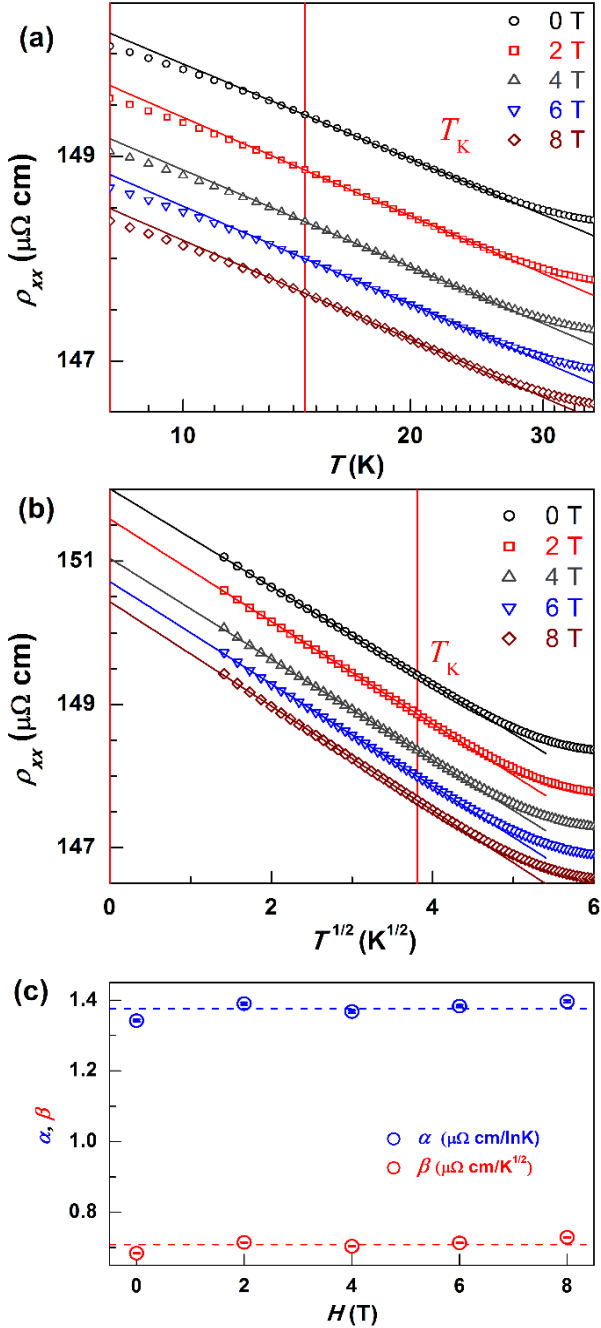

Fig. 7. (a) Semilog plot of $\rho_{xx}$ versus $T$ and (b) $\rho_{xx}$ versus $T^{1/2}$; (c) $H$ dependence of $\alpha$ and $\beta$ for the $L1_0$-MnGa film. For clarity, $\rho_{xx}$ was shifted by 0, -0.2, -0.4, -0.6, and -0.8 $\mu\Omega$ cm in (a) and (b), respectively. The Kondo temperature $T_K$ is 14.5 ±1.5 K. The dashed lines in (c) are for eye guidance.

### 2.3 Coexistence of the orbital 2CK effect with ferromagnetism

The evident coexistence of the 2CK physics and ferromagnetism is an intriguing observation. Although the two spin channels are still degenerate in energy because the Kondo coupling with a TLS is nonmagnetic and does not involve any spin variables, the population imbalance of the two spin channels due to the ferromagnetic exchange splitting of the conduction band could be significant in comparison to the magnetic field effects for the TLS model. In fully ordered $L1_0$-MnGa and $L1_0$-MnAl, the spin moments of Mn atoms are parallel due to ferromagnetic Ruderman–Kittel–Kasuya–Yoshida (RKKY) interaction with the spin polarization being dominantly determined by the Mn $3d$ states.[22,23] In disordered samples, the Mn-Mn antiparallel alignment due to antiferromagnetic superexchange simultaneously reduces $M_s$ and spin polarization as a consequence of the compensating contributions from the oppositely aligned Mn atoms.[1,8,23,24] The spin polarization should be proportional to $M_s$ if the valence electron number ($N_v$) is assumed to independent of the disorder ($N_v$ is 10 for both MnGa and MnAl,[25] respectively). For the disordered $L1_0$-MnGa and MnAl films showing an orbital 2CK effect, the value of $M_s$ is only 12.5% (10%-40%) of the theoretical value for the fully ordered $L1_0$-MnGa (MnAl),[16,17] indicating a robust antiparallel alignment of Mn-Mn magnetic moments and a very low degree of spin population imbalance. This could be the reason why the ferromagnetism does not quench the 2CK physics here. The robust 2CK effect observed in ferromagnetic systems, e.g. $L1_0$-MnGa and $L1_0$-MnAl, also hints that the fixed point of an orbital 2CK effect is more robust to the loss of spin population balance in comparison to that of a spin 2CK effect to the orbital channel asymmetry.[16,17] However, a dilution of NFL behavior and an enhancement of $T_D$ due to the loss of spin population balance is expected in a ferromagnet with a partially spin-polarized conduction band. It would be very interesting to quantitatively determine how the stability of 2CK fixed point varies with an enhancing population imbalance of the spin channels. For example, the spin polarization of the host materials may be manipulated either by engineering the film structure quality via growth parameters[26] or by injecting a pure spin current from an adjacent spin Hall generator, e.g. heavy metals (Ta, W or Pt)[27-29] or antiferromagnet (IrMn or PtMn).[30,31] More theoretical and experimental efforts are needed to better understand the exotic 2CK physics, especially in ferromagnetic hosts.

### 3. Anomalous Hall effect

As shown in Fig. 8(a), the AHE occurs when a charge current flows through a conductor with a magnetic moment along the normal direction.[17] The AHE is now widely accepted to include the intrinsic deflection due to Berry curvature of Bloch states and the extrinsic skew scattering and side jump resulting from the spin-orbit-interaction–induced asymmetrical scattering of conduction electrons (Fig. 8(b)).[32-34] Skew scattering yields a scaling relation between $\rho_{xx}$ and anomalous Hall resistivity ($\rho_{AH}$) as $\rho_{AH} \sim \rho_{xx}$, while the other two mechanisms give $\rho_{AH} \sim \rho_{xx}^2$. Accordingly, scaling laws $\rho_{AH} \sim \rho_{xx}^n$ and $\rho_{AH} = a\rho_{xx} + b\rho_{xx}^2$ are routinely used to describe the experimental data and decipher the relevant mechanisms for the AHE. In contrast, some recent studies in Fe and Co films revealed a negligible contribution of phonon skew scattering and a scaling of $\rho_{AH}=a_0\rho_{xx0}+b\rho_{xx}^2$,[35,36] where $\rho_{xx0}$ is the residual resistivity induced by impurity scattering and $a_0\rho_{xx0}$ is the extrinsic contributions of skew scattering and side jump from impurities. These intriguing observations make the AHE scaling open question.

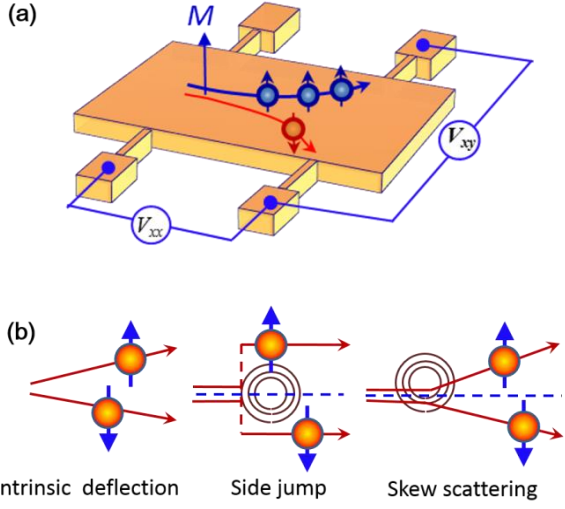

Fig. 8. (a) Schematic of the AHE and measurement configuration; (b) the AHE mechanisms: intrinsic deflection, side jump, and skew scattering.

Alloys of $L1_0$-MnGa and MnAl are ideal for a systematic examination of the AHE scaling and the underlying physics, as their perfectly square hysteresis allow for highly accurate determination of $\rho_{AH}$, large magnitude of $\rho_{AH}$ (e.g. 2-7.5 μΩ cm in $Mn_{1.5}Al$ films grown on AlAs-buffered GaAs),[37,38] and their degree of structural ordering could be controllably varied for tailoring the related transport behaviors.

### 3.1 AHE in high-quality $L1_0$-MnGa films

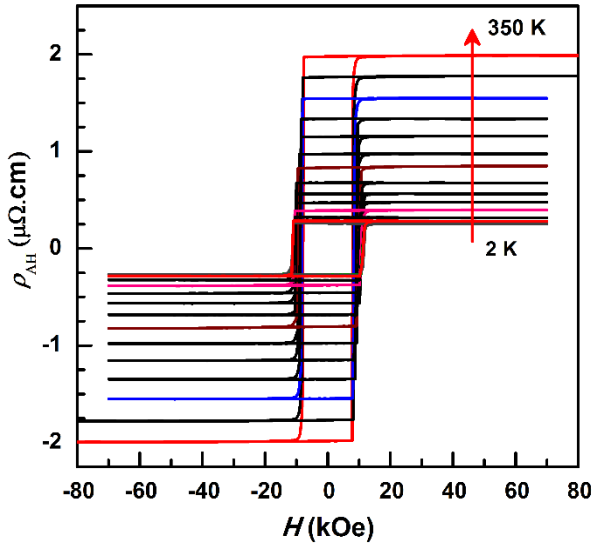

Fig. 9. $T$-dependent hysteretic anomalous Hall resistivity of an $L1_0$-MnGa film grown at 250 °C.

So far, the AHE of $L1_0$-MnGa films has been reported by several groups. Tanaka et al. and Zhu et al. studied AHE of $L1_0$-MnGa ($x = 1.5$) films on GaAs and found $\rho_{AH}$ to vary from 0.3-4 μΩ cm depending on the samples and $T$.[39,40] Bedoya-Pinto et al. reported composition-dependent $\rho_{AH}$ in $L1_0$-MnGa ($x$=0.96, 1.38, and 2.03) films on GaN.[41] Recently, the scaling behaviors of the AHE were systematically studied in a series of 50 nm thick $L1_0$-MnGa ($x$=1.5) single-crystalline films epitaxially grown on GaAs (001) by molecular-beam epitaxy.[40] Figure 9 shows an example of the fairly square $\rho_{AH}$-$H$ curves of the $L1_0$-MnGa epitaxial films at different $T$ ranging from 2 to 350 K, suggesting the high structural quality and giant perpendicular anisotropy in these films. Detailed scaling analysis clarifies that $\rho_{AH}=a_0\rho_{xx0}+b\rho_{xx}^2$ is the only correct scaling law for the materials where defect scattering and phonon scattering dominate the electron scattering. Figure 10 plots $\rho_{AH}$ as a function of $\rho_{xx}^2$ for the $L1_0$-MnGa films at 100, 150, 200, 250 and 300 °C, from which the excellent agreement between the scaling law and experimental data can be seen. The conventional scaling laws $\rho_{AH}\sim\rho_{xx}^n$ and $\rho_{AH}=a\rho_{xx}+b\rho_{xx}^2$ are found to evidently deviate from the experimental data.

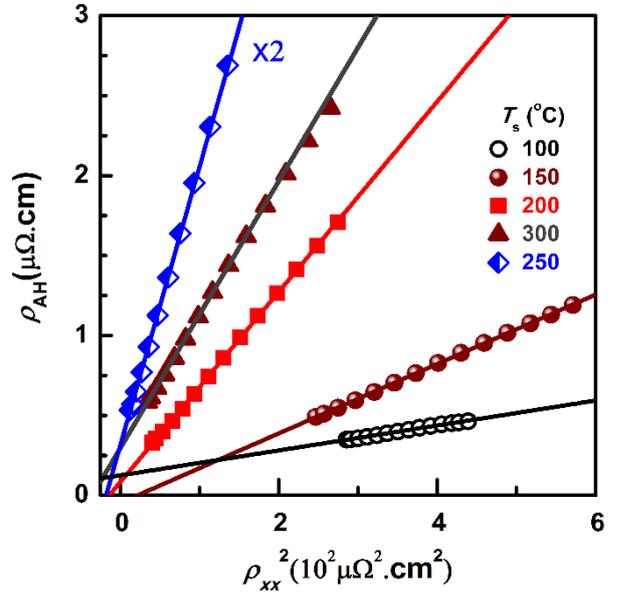

Fig. 10. $\rho_{xx}^2$ dependence of $\rho_{AH}$. The lines are fits to $\rho_{AH}=a_0\rho_{xx0}+b\rho_{xx}^2$ for the data at different temperatures. For better comparison, the values of $\rho_{AH}$ for $T_s$ = 250 °C (blue) are multiplied by a factor of 2.

### 3.2 AHE in $L1_0$-MnAl films with the orbital 2CK effect

As discussed above, the scaling law $\rho_{AH}=a_0\rho_{xx0}+b\rho_{xx}^2$ excellently describes the AHE for the materials where defect scattering and phonon scattering dominate the electron scattering. However, in the presence of strong disorder effects (e.g. hopping conduction, weak localization, and electron-electron interaction),[42-45] the AHE scaling is more complex and under debate. In the orbital 2CK effect where the TLSs, the localized "pseudo-spin" 1/2 impurities, strongly and equally couple to the conduction electrons from two spin channels, the AHE scaling has been unclear. The recent observation of the robust and controllable orbital 2CK effect in $L1_0$-MnAl ferromagnetic films,[16] provide an experimental access to the AHE in the presence of the orbital 2CK effect.[46]

Figure 11 show the scaling behavior of the AHE for a series of 30 nm-thick $L1_0$-MnAl films grown at 200, 250, 300, 350 and 400 °C, respectively. The AHE scaling is observed to follow $\rho_{AH}/f=a_0\rho_{xx0}+b\rho_{xx}^2$ at high $T$ where phonon scattering prevails.[46] Here $f = M / M_0$, where $M$ ($M_0$) is magnetization at finite $T$ (0 K); the $T$ independence of $a_0$ and $b$ is not considered. $f$ is a correction to $\rho_{AH}$ because of the relatively low $T_c$ and resultant $T$ dependence of $M$. In contrast, the AHE significantly deviates from it at low $T$ where the orbital 2CK effect becomes important. The breakdown of the scaling seems closely correlated to the orbital 2CK effect. The breakdown temperatures of the scaling and the magnitudes of the deviation at 2 K for different samples are excellently consistent with the trig-on temperature and the strength of the orbital 2CK effect, which is highly reminiscent of a close relation between this breakdown and the orbital 2CK effect.

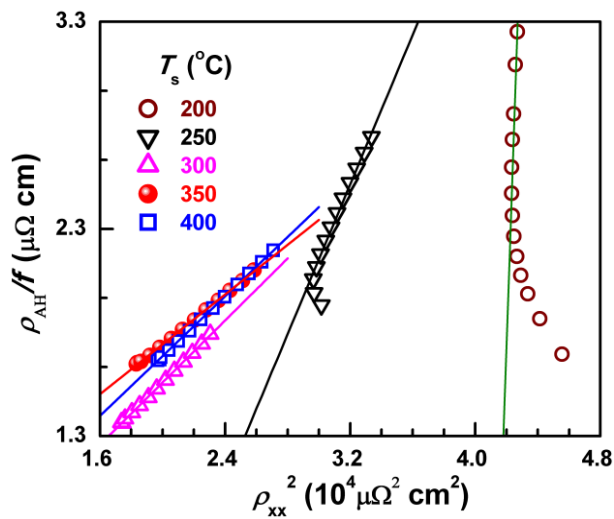

Fig. 11. $\rho_{AH}/f$ vs $\rho_{xx}^2$ for $L1_0$-MnAl films grown at 200, 250, 300, 350, and 400 ºC.

## 4. Summary and outlook

Magnetic films with giant PMA and controllable epitaxial growth are potentially useful as the platform for further exploration of novel fundamental physics. In this article, we discussed the recent observations of the strongly correlated physics of the orbital 2CK effect and the underlying physics of the AHE taking advantage of the $L1_0$-ordered MnGa and MnAl films with giant PMA. The controllable structural, magnetic, and electronic disorders give insights into the physical mechanisms of disorder-related transport phenomena.

Besides, the $L1_0$-ordered MnGa and MnAl alloys have many potential directions to benefit the scientific and technologic progress. For example, $L1_0$-ordered MnGa and MnAl alloys show promise for future perpendicular magnetic recording with areal density beyond 10 Tb/inch$^2$ that requires high PMA of up to ~10 Merg/cc and moderate magnetization $M_s$. These materials are also potentially useful as economical permanent magnets applications due to their rare-earth-free and noble-metal-free composition, large magnetic energy product, high coercivity and linear demagnetization curve in the second quadruple.[47] Especially, future efforts on these high-PMA materials may greatly benefit the nanoscale spintronic applications, e.g. spin valves and magnetic tunneling junctions based magnetoresistive random access memory and oscillators driven by current-induced spin transfer torque (STT) or spin orbit torque (SOT). A PMA free layer with relatively low $M_s$ and small $\alpha$ allow to facilitate STT and SOT switching of MRAMs and to excite oscillators with a low critical current density. The ultrahigh coercivity, e.g. ~4.3 T at room temperature in some cases,[4] makes them fascinating as the perpendicular reference layer of an orthogonal spin valve or magnetic tunneling junction that possess an inplane magnetized free layer and can be used as linear high-magnetic-field sensors.[48] $L1_0$-ordered MnGa and MnAl alloys also show good compatibility with semiconductor, which could benefit the development of room temperature ferromagnetic semiconductors via magnetic proximity effect and ferromagnetic metal/semiconductor hybrid devices with perpendicular anisotropy.[49,50] In addition, as a consequence of the high PMA, the moderate polar magnetooptical Kerr angles, and high reflectivity,[51,52] $L1_0$-MnGa and $L1_0$-MnAl films are also interesting in the spatially-resolved optical studies of spintronic phenomena, such as the magnitudes and the symmetries of SOTs,[53] the spin-torque generation and manipulation of skyrmions,[54] the magnetic domain wall motion,[55] and the magnetization switching driven by a charge current flowing in an adjacent spin Hall layer. High-quality $L1_0$-MnGa films have been shown to exhibit spin precession at terahertz frequency range due to their giant PMA,[56] which would be an interesting candidate material to study terahertz spin pumping and spin injection.

Keeping in mind the amazing properties of the materials, we expect more detailed explorations of the practical applications and fundamental physics of $L1_0$-MnGa and MnAl in the fields of magnetism, spintronics, and strongly correlated phenomena in the near future.

## Acknowledgements

The work was supported partly by MOST of China (grant no. 2015CB921503), NSFC (grant nos. 61334006, U1632264) and the Key Research Project of Frontier Science of CAS [grant no. QYZDY-SSW-JSC015].